\documentclass[%
 reprint,
superscriptaddress,
 amsmath,amssymb,
 aps,
pra,
]{revtex4-2}

\usepackage{graphicx}
\usepackage{dcolumn}
\usepackage{bm}
\usepackage{float}
\usepackage{hyperref}
\usepackage{braket}
\usepackage{physics}
\usepackage{xcolor}
\usepackage{soul}
\usepackage[normalem]{ulem}
\usepackage{upgreek}
\usepackage{soul}
\usepackage{amssymb}
\usepackage{multirow}
\usepackage{enumitem,kantlipsum}

\begin{document}

\preprint{APS/123-QED}

\title{A Generalized framework of weak value amplification  in path interference of polarized light for  the enhancement of all possible polarization anisotropy effects} 

\author{Niladri Modak}
\email{nm16ip018@iiserkol.ac.in}
\affiliation{Department of Physical Sciences, Indian Institute of Science Education and Research (IISER) Kolkata. Mohanpur 741246, India.}
\author{Athira B S}
\affiliation{Center of Excellence in Space Sciences India, Indian Institute of Science Education and Research (IISER) Kolkata. Mohanpur 741246, India.}


\author{Ankit kumar Singh}
\affiliation{Department of Physical Sciences, Indian Institute of Science Education and Research (IISER) Kolkata. Mohanpur 741246, India.}
\author{Nirmalya Ghosh}
\affiliation{Department of Physical Sciences, Indian Institute of Science Education and Research (IISER) Kolkata. Mohanpur 741246, India.}
\affiliation{Center of Excellence in Space Sciences India, Indian Institute of Science Education and Research (IISER) Kolkata. Mohanpur 741246, India.}


\date{\today}

\begin{abstract}
Using the profound interferometric philosophy of weak value amplification, we propose a simple, general and a robust polarization method for the amplification and quantification of small magnitudes of all possible polarization anisotropy effects in a single experimental embodiment.
 The approach is experimentally realized by introducing a weak coupling between the polarization degree of freedom of light and the path degree of freedom in a Mach-Zehnder interferometer in the presence of a weak anisotropy effect. Real and imaginary weak value amplifications of different polarization anisotropy effects are manifested as characteristic changes in relevant Stokes vector elements at the exit port of the interferometer, which follow orthogonal trajectories in the Poincare sphere.  
The proof of concept experiment demonstrates that using this scheme, one can faithfully extract and quantify anisotropy parameter that is smaller than the typical sensitivity of measurement of a given Stokes parameter of a traditional polarimeter by a large weak value amplification factor.
This opens up possibility of a sample measuring \textit{weak value polarimeter} for studying rich variety of fundamental optical effects and for materials characterization and precision metrology.
\end{abstract}
\maketitle
\section{\label{sec:level1}Introduction}
Polarized light measurements have played vital role in understanding and developing various advanced concepts of electromagnetic waves , and in numerous applications in diverse fields ranging from astronomy, materials characterization, biomedical imaging, remote sensing to meteorology \cite{bliokh2008geometrodynamics,bliokh2015quantum,aiello2015transverse,lin2014dielectric,pal2019experimental,perez2016polarized,azzam2016stokes}. Polarimetry techniques probe anisotropic polarizability of matter through the so-called birefringence (retardance) and dichroism (diattenuation) parameters \cite{collett1993polarized,gupta2015wave,perez2016polarized,azzam2016stokes}. These are traditionally quantified through exhaustive measurement of the $4\times4$ sample Mueller matrix or by multiple measurements of the $1\times4$ Stokes vector elements of sample-emerging light \cite{collett1993polarized,gupta2015wave,perez2016polarized,azzam2016stokes}. These conventional polarimetry techniques are however, compromised when one needs to measure extremely small polarization effect desirable for certain applications, e.g., for the detection of physiological glucose concentration in human tissue, for the quantification of extremely small circular dichroism of proteins, for the determination of weak magneto-optical rotation in nanomaterials and so on \cite{collett1993polarized,gupta2015wave,perez2016polarized,azzam2016stokes}. Measurements employing polarization modulation and synchronous detection have thus been developed \cite{collett1993polarized,gupta2015wave,perez2016polarized,azzam2016stokes}. Despite the availability of different such advanced techniques, there is a renewed recent interest in developing polarization methods that are particularly founded on rich fundamental principle of wave optics, are technically simple, can obviate tedious measurements and calibration procedure, and yet are capable of providing high accuracy and sensitivity. Such methods are of both fundamental and applied interests, e.g., for studying rich variety of spin (polarization) optical effects and various intriguing quantum optical phenomena, for the characterization of weakly anisotropic materials and for precision metrology \cite{bliokh2015quantum,lin2014dielectric,pal2019experimental,athira2020single,perez2016polarized,azzam2016stokes,collett1993polarized,gupta2015wave,aharonov1988result,pal2019experimental,singh2018transverse}.
\par
Despite of being developed in the field of quantum mechanics \cite{aharonov1988result,duck1989sense,kofman2012nonperturbative,dressel2014colloquium}, in recent times, weak measurements and weak value amplification (WVA) have proven to be fundamentally important and extremely useful in the realm of both classical and quantum physics \cite{ritchie1991realization,hosten2008observation,magana2014amplification,xu2013phase,dixon2009ultrasensitive,asano2016anomalous,PhysRevA.89.053816,nechayev2018weak,shomroni2013demonstration,singh2018tunable,lundeen2011direct,palacios2010experimental,kocsis2011observing,salvail2013full,luo2019simultaneously,xu2019multifunctional,pryde2005measurement,de2014ultrasmall} due to its origin in wave interference \cite{duck1989sense,ritchie1991realization,dressel2014colloquium,kofman2012nonperturbative}. The WVA mechanism using post-selected weak measurements \cite{ferrie2014weak} has attracted particular attention in the optical domain for addressing fundamental questions \cite{pal2019experimental,magana2014amplification,PhysRevA.89.053816,palacios2010experimental,kocsis2011observing,salvail2013full,berry2011pointer,zhu2019single} as well as for potential applications \cite{magana2014amplification,xu2013phase,dixon2009ultrasensitive,shomroni2013demonstration,singh2018tunable,salvail2013full,luo2019simultaneously,xu2019multifunctional,pryde2005measurement,de2014ultrasmall}. The WVA protocol sequentially involves the preparation of a system state (pre-selection), a weak coupling between the system and a pointer, and a post-selection \cite{aharonov1988result,duck1989sense,kofman2012nonperturbative,dressel2014colloquium}. Near mutual orthogonal pre- and post-selection of states gives rise to  near destructive interference between the slightly separated pointer profiles leading to a large deflection of the resultant pointer profile, which is interpreted as the WVA of an observable \cite{aharonov1988result,duck1989sense,kofman2012nonperturbative,dressel2014colloquium}. WVA has been successfully used in classical optics for numerous practical applications, e.g., to amplify  and detect tiny optical beam deflection \cite{dixon2009ultrasensitive} and spin Hall shift of light \cite{hosten2008observation,pal2019experimental}, for sensitive estimation of angular rotation \cite{magana2014amplification}, for high resolution phase and frequency measurements \cite{xu2013phase,luo2019simultaneously}, for the measurement of ultra-small time delays etc. 
Weak measurement has also been widely explored in the quantum optics domain  \cite{lundeen2011direct, kocsis2011observing, palacios2010experimental, PhysRevLett.109.040401}. 
\par In the context of polarization measurements, interesting weak measurement protocols have been developed for the determination of the quantum weak values of single photon's polarization \cite{pryde2005measurement,iinuma2011weak}, for direct measurement of general polarization state of light \cite{salvail2013full}, or to perform polarization state tomography of light field \cite{zhu2019single}. These measurement schemes are aimed at finding out the weak values of the Stokes polarization operators of single photon or classical light beams \cite{pryde2005measurement,iinuma2011weak}. 
While several weak measurement protocols have been developed for the characterization of the polarization state of light, relatively lesser amount of research has been done for the extraction and quantification of small polarization anisotropy properties of sample \cite{xu2019multifunctional,luo2019simultaneously,de2014ultrasmall,rhee2015chiroptical}. In a recent study, optical activity of a sample was probed by high precision measurement of the changes in the amplitude and the phase of a light beam via measurement of both the real and the imaginary (respectively) weak values using different polarization state post selections \cite{luo2019simultaneously}. In another weak measurement protocol, the optical rotation signal of a chiral sample was 
amplified and detected by introducing a weak coupling between the polarization and the spectral degree of freedom of light and by subsequent pre and post-selection of states using appropriate polarizations \cite{xu2019multifunctional}. Similarly, the optical polarization rotation was successfully amplified by adopting the angular version of von Neumann measurement scheme \cite{de2014ultrasmall}. 
Another useful technique that worth a mention here is the so-called quasi-null-polarization-detection method to amplify small optical rotatory dispersion signal from chiral samples \cite{rhee2015chiroptical}. Even though this is not a WVA scheme per say, in this method also, the chiral sample is illuminated with weakly elliptically polarized light and subsequent post selection is done in near orthogonal linear polarization. This essentially results in a large increase of the ratio of the chiral to the achiral signal intensity, leading to amplification of the optical rotatory dispersion signal. \par
In this paper, we introduce 
a polarization measurement technique capable of amplifying and quantifying small magnitudes of all the polarization anisotropy effects of sample based on interferometric philosophy of post-selected weak measurement \cite{aharonov1988result,duck1989sense}.  
In this approach, near destructive interference of two paths in an interferometer serves the purpose of near orthogonal pre-post selection  of states \cite{dixon2009ultrasensitive} unlike  conventional optical schemes of WVA. A weak polarization anisotropy effect introduced in one path of the interferometer provides the desirable weak coupling between the path degree of freedom and the polarization degree of freedom of light. The real and the imaginary WVA of different anisotropy effects are manifested in different characteristic Stokes vector elements (acting as the pointer here) \cite{collett1993polarized,gupta2015wave} at the exit port of the interferometer enabling quantification of all the anisotropy effects in the same experimental embodiment without involving any additional anisotropy effect-dependent post-selections. Note that interferometric arrangement have also been used previously for obtaining weak value amplification \cite{dixon2009ultrasensitive, dziewior2019universality}. However, in most of those schemes the spatial degree of freedom of light (spatial mode) has been used as a pointer whereas the paths of the interferometric setup have been given two different polarization. The WVA in such scenario manifests as large change in the pointer beam profile either in the spatial or in the momentum space. In contrast, our scheme incorporates the polarization state of light itself as a pointer enabling WVA of polarization anisotropy effects. To the best of our knowledge, such an interferometric WVA scheme for WVA of all possible  polarization anisotropy effects in a single experimental embodiment has not been reported previously.
\par 
The paper is organized as follows. In Section.\ref{sec:level2}, we provide  
the theoretical framework for interferometric WVA of all the polarization anisotropy effects along with its corresponding classical field-based formalism. 
Section.\ref{sec:level3} provides  the experimental details of interferometric WVA. In Section.\ref{sec:level4}, the experimental and simulation results 
are presented and discussed. Section.\ref{sec:level5} concludes with an outlook on this new type of weak value polarimeter, its possible extension towards imaging and spectroscopic polarimetry and its potential applications.
\begin{figure*}
\includegraphics[width=150mm]{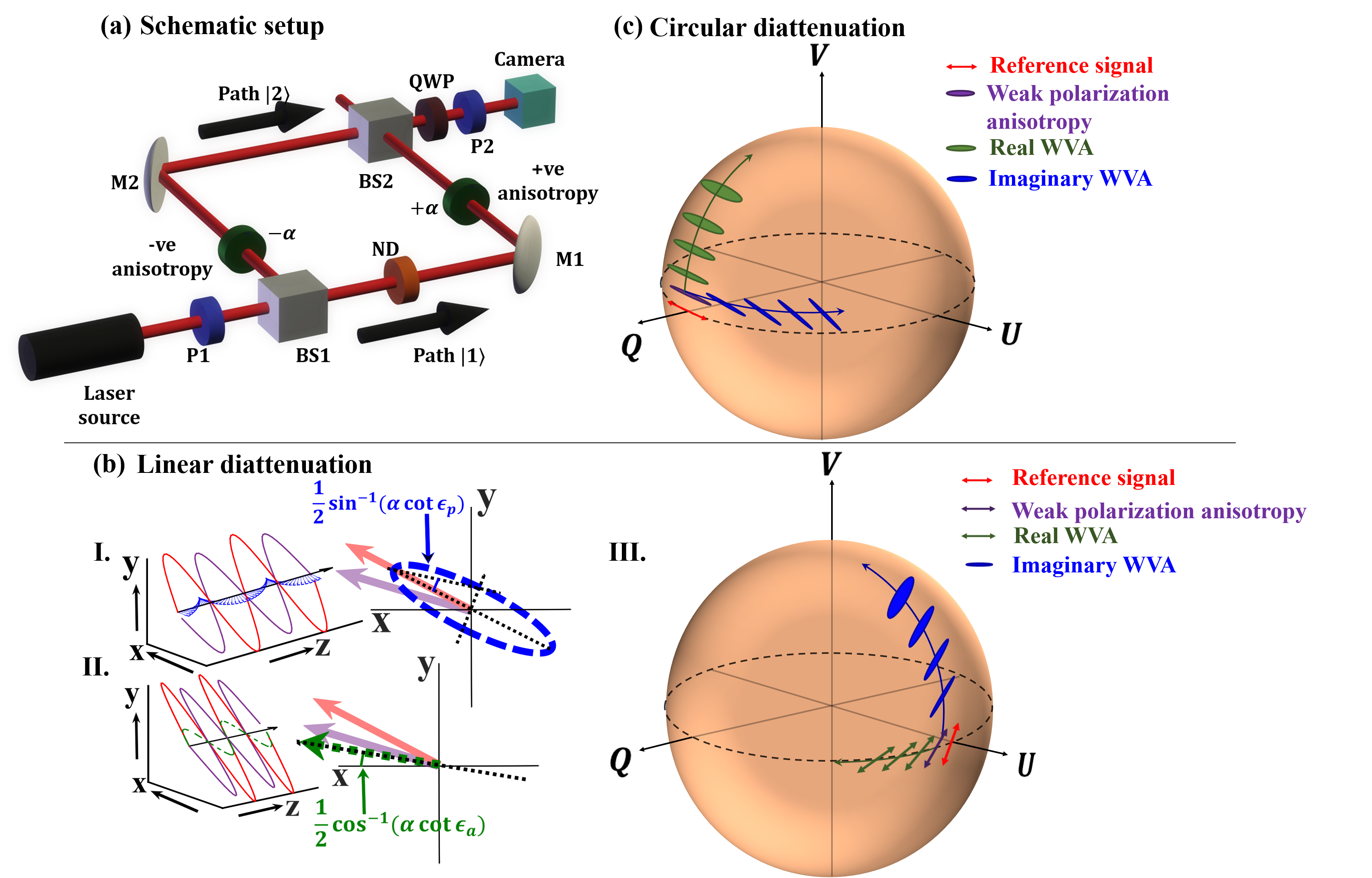}
\caption{(a) Schematic of the Mach-Zehnder interferometric arrangement for WVA of polarization anisotropy effects. (P1,P2): polarizers, QWP: quarter waveplate, (BS1, BS2): 50:50 beam splitters, (M1, M2): mirrors, ND: variable neutral density filter. WVA of (b) \textit{linear diattenuation} and (c) \textit{circular diattenuation}. (b) Electric field vectors and polarization ellipse undergoing \textit{imaginary} (\textbf{\textit{I}}) and \textit{real} (\textbf{\textit{II}}) WVA are illustrated. \textit{\textbf{I:}} Electric fields corresponding to the $+45^{o}$ polarized light in path 1 (red solid line), after experiencing a small linear diattenuation in path 2 (violet solid line) and the resultant field following near destructive interference with small phase offset $\epsilon_p$ (blue dashed line) are depicted. Enhancement of ellipticity due to \textit{imaginary} WVA is also displayed and the  expression for ellipticity is noted. \textit{\textbf{II:}} Corresponding electric fields for \textit{real} WVA.  The resultant field after near destructive interference with small amplitude offset $\epsilon_a$ is shown (green dashed line). The change in the orientation angle of the electric field due to real WVA is also displayed and the corresponding expression is noted. \textit{\textbf{III:}} Evolution of the Stokes vectors in the Poincare sphere for real (green line) and imaginary (blue line) WVA of linear diattenuation. (c) Evolution of the Stokes vectors in the Poincare sphere for real (green line) and imaginary (blue line) WVA of circular diattenuation.}
\label{figure1}
\end{figure*}
\section{\label{sec:level2}THEORETICAL FORMALISM OF INTERFEROMETRIC WVA OF POLARIZATION ANISOTROPY EFFECTS}
In this framework, the weak value of a given polarization anisotropy is realized through near destructive interference of polarized light in an interferometer (see Fig.\ref{figure1}a). Here, near destructive interference refers to destructive interference of two fields having a phase difference $\pi$ with a small amplitude offset $\epsilon_a=\tan^{-1}({\frac{1-a}{1+a}})$ ($a$ is the ratio of amplitudes of the two fields) or interference of two fields having equal amplitude ($a = 1$)  but with phase difference of $\pi\pm2\epsilon_p$ where $\epsilon_p$ is small phase offset. The former generates \textit{real} WVA and the latter \textit{imaginary} WVA. The $\epsilon_{a\slash p}$ parameters are equivalent to the overlap of near orthogonal pre-post selected states in weak measurements. We define the generalized polarization anisotropy parameter $\alpha$ in terms of the difference in refractive index, RI ($n$) between orthogonal linear or circular polarizations as, $2\alpha=\frac{2\pi}{\lambda} (n_{L\slash C}^{r\slash i}-n_{L'\slash C'}^{r\slash i})t$. Here, $t$ is the path length, and the superscripts ($r\slash i$) correspond to real/imaginary parts of RI, and the subscripts ($L\slash C$) and ($L'\slash C'$ ) represent linear/circular polarizations and their orthogonal states, respectively. The real/ imaginary parts of RI generate amplitude/ phase anisotropies and are associated with retardance/ diattenuation effects, respectively \cite{collett1993polarized,gupta2015wave}. We now derive an interferometric WVA framework based on the post-selection aided weak measurement formalism.
\subsection{\label{sec:level2a}Post-selected weak measurement formalism}
The weak coupling between the path degree of freedom (system) and polarization state (pointer) of light beam in the interferometric arrangement (see Fig.\ref{figure1}a) can be expressed by the unitary evolution $U(\alpha)=e^{-i \alpha G \hat{A}}$ \cite{von2018mathematical,aharonov1988result,duck1989sense}. Here the polarization anisotropy parameter $\alpha\rightarrow 0$, $G$ is the generator of the polarization effect acting as the poniter variable \cite{han1997stokes}. It generates $2\times2$ Jones matrices corresponding to the different polarization anisotropy effect. 
Henceforth, we use the superscript $\mathfrak{a}\slash\mathfrak{p}$ to describe amplitude/phase anisotropy effect in the corresponding generators $G_j^\mathfrak{a}=\frac{i}{2}\sigma_j\ \text{and} \ G_j^\mathfrak{p}=\frac{1}{2}\sigma_j$ respectively, $\sigma_j\ (j=1,2,3)$ are  the standard $2\times 2$ Pauli matrices. In our context, the subscripts $j=1,2,3$ are related to the generator for the Jones matrix corresponding to the anisotropy effects in the $\pm45^{o}$ linear polarization, right- left circular polarization, and horizontal ($x$)- vertical ($y$) linear polarization basis respectively \cite{han1997stokes}. The observable $\hat{A}$ representing opposite polarization anisotropy effect $+\alpha\ \text{and}\ -\alpha$ in the two paths (path $\ket{1}$ and path $\ket{2}$ respectively) of the interferometer can be expressed as a $2\times2$ diagonal matrix
    $\hat{A}=\ket{1}\bra{1}-\ket{2}\bra{2}=
    \begin{bmatrix}
    1 & 0\\
    0 & -1
    \end{bmatrix}$ \cite{dixon2009ultrasensitive}.
Pre-selection ($\ket{\psi_i}$) and post-selection ($\ket{\psi_f}$) of states can be obtained by path degree of freedom as follows,
\begin{subequations}
\begin{equation}
    \ket{\psi_i}=\frac{1}{\sqrt{2}}[\ket{1}+\ket{2}]
    \label{eq3a}
\end{equation}
 \begin{equation}
   \ket{\psi_f}=\frac{1}{\sqrt{1+a^2}}[e^{\pm i\epsilon_p}\ket{1}-a e^{\mp i\epsilon_p}\ket{2}]\label{eq3b}
   \end{equation}
   \label{eq2}
\end{subequations}
\noindent 
In the weak coupling limit ($\alpha \rightarrow 0$), the final pointer state after post selection is obtained as \cite{kofman2012nonperturbative,dressel2014colloquium,aharonov1988result,duck1989sense},
    \begin{equation}
    \ket{\phi_f}\approx\braket{\psi_f}{\psi_i}e^{-i\alpha G A_w}\ket{\phi_i}
    \label{eq3}
\end{equation}
Here $A_w=\frac{\mel{\psi_f}{\hat{A}}{\psi_i}}{\braket{\psi_f}{\psi_i}}$ is the interferometric weak value of the operator $\hat{A}$. $\ket{\phi_i}$ is the initial pointer state, which is the input polarization state represented by Jones vector. 
In this framework, the polarization transformation is determined by Eq.\eqref{eq3} and accordingly the final state will depend upon the values of $A_w\ \text{and}\ \alpha$.
\subsubsection{\textit{Real weak value amplification}}
Real WVA is obtained at the exact destructive interference position (phase difference of pi, $\epsilon_p=0$) with small amplitude offset $\epsilon_a$. Using Eq.\eqref{eq2} and \eqref{eq3}, the final pointer state after post-selection is obtained as,
\begin{equation}
    \ket{\phi_f}\sim e^{-i\alpha\cot{\epsilon_a}G}\ket{\phi_i}
    \label{eq4}
\end{equation}
Eq.\eqref{eq4} implies an enhancement of the polarization effect $\alpha$ by a factor $\cot{\epsilon_a}$ , which is reflected in the final Jones vector pointer state.

\subsubsection{\textit{Imaginary weak value amplification}}
Imaginary WVA is obtained by near destructive interference of two paths with equal amplitude ($a=1$) but with a phase offset $\epsilon_p$ from exact destructive interference position. The final pointer state after post-selection, in this case, becomes
\begin{equation}
    \ket{\phi_f}\sim e^{\alpha\cot{\epsilon_p}G}\ket{\phi_i}
    \label{eq5}
\end{equation}
Depending upon the anisotropy effect (described by generator $G$), the Jones vector pointer state will be characteristically modified.
\par
For example, in the case of optical rotation $\alpha$, the pointer variable is expressed by $G_2^\mathfrak{p}$ with input $x$-polarized state (Jones vector) as $\ket{\phi_i}=\begin{bmatrix}
    1&0
    \end{bmatrix}^T$. Following Eq.\eqref{eq4}, the real WVA is obtained in the final pointer state $\ket{\phi_f}$.
    \begin{equation}
    \ket{\phi_f}\sim e^{-i\alpha\cot{\epsilon_a}G_2^\mathfrak{p}}\ket{\phi_i}\sim \begin{bmatrix}\cos({\alpha\cot{\epsilon_a}})\\ -\sin({\alpha\cot{\epsilon_a}})
\end{bmatrix}
    \label{eq6}
    \end{equation}
Clearly, from Eq.\eqref{eq6}, the real WVA of small optical rotation $\alpha$ is manifested as a large enhancement in the orientation angle ($=\alpha\cot{\epsilon_a}$) of the output Jones vector \cite{gupta2015wave}. This is valid for the case $\epsilon_a\rightarrow0$.
\par
Similarly, using Eq.\eqref{eq5}, the final state after post-selection for an imaginary WVA of small optical rotation can be obtained as,
\begin{equation}
\ket{\phi_f}\sim e^{i\alpha\cot{\epsilon_p}G_2^\mathfrak{a}}\ket{\phi_i}\sim \begin{bmatrix}\cosh({\alpha\cot{\epsilon_p}})\\ -i\sinh({\alpha\cot{\epsilon_p}})
\end{bmatrix}
\label{eq7}
\end{equation}
Imaginary WVA, in Eq.\eqref{eq7}, is manifested as a large change in the ellipticity of the output Jones vector \cite{gupta2015wave} with decreasing  $\epsilon_p$.
\begin{table*}[]
\caption{Real and imaginary WVA of all the polarization anisotropy effects. The four different polarization anisotropy effects (1st column), the corresponding input electric field and the electric field after experiencing a small anisotropy effect (2nd column), the Stokes vector elements carrying signature of real and imaginary WVA of respective anisotropy effects (3rd column). The polarization optical components used to experimentally realize the WVA of respective anisotropy effects are listed in the 4th column.}
\label{table1}
\begin{ruledtabular}
\begin{tabular}{lccccr}
\centering
\multirow{2}{*}{Anisotropy effects} & \multicolumn{2}{c}{Electric fields}                                                                   & \multicolumn{2}{c}{Stokes vector elements}& \begin{tabular}[c]{@{}r@{}}Polarization\\ optical component    \end{tabular}                                                                                        \\\\
                                    & \begin{tabular}[c]{@{}c@{}}$\frac{\pmb{E_1}}{\mathfrak{\xi}}$ \end{tabular}                       & $\frac{\pmb{E_2}}{\mathfrak{\xi}}$                                                           & Real WVA                       & Imaginary WVA                  &                                          \\                                                                 \colrule                                 \\
Linear diattenuation\\ ($x -y$)       & \begin{tabular}[c]{@{}c@{}}$\frac{\hat{x}+\hat{y}}{\sqrt{2}}$\\ $+45^{o}$ linear polarization\end{tabular} & $\frac{e^\alpha\hat{x}+e^{-\alpha}\hat{y}}{\sqrt{e^{2\alpha}+e^{-2\alpha}}}$ & $\frac{Q}{I}$ & $\frac{V}{I}$ & \begin{tabular}[c]{@{}r@{}}Linear polarizer\\ oriented at a\\ small angle from $+45^{o}$\end{tabular}                                       \\\\
Circular diattenuation              & \begin{tabular}[c]{@{}c@{}}$\hat{x}=\frac{\hat{r}+\hat{l}}{\sqrt{2}}$\\ Horizontal linear\\ polarization\end{tabular} & $\frac{e^\alpha\hat{r}+e^{-\alpha}\hat{l}}{\sqrt{e^{2\alpha}+e^{-2\alpha}}}$ & $\frac{V}{I}$ & $\frac{U}{I}$ & \begin{tabular}[c]{@{}r@{}}Quarter waveplate with its\\ optic axis oriented at\\ a small angle from\\ horizontal polarization\end{tabular} \\\\
Linear retardance                   & \begin{tabular}[c]{@{}c@{}}$\frac{\hat{x}+\hat{y}}{\sqrt{2}}$\\ $+45^{o}$ linear polarization\end{tabular} & $\frac{e^{i\alpha}\hat{x}+e^{-i\alpha}\hat{y}}{\sqrt{2}}$             & $\frac{V}{I}$ & $\frac{Q}{I}$ & \begin{tabular}[c]{@{}r@{}}Liquid crystal variable retarder \\ with small retardance\end{tabular}                                          \\\\
Optical rotation                    & \begin{tabular}[c]{@{}c@{}}$\hat{x}=\frac{\hat{r}+\hat{l}}{\sqrt{2}}$\\ Horizontal linear\\ polarization\end{tabular} & $\cos{\alpha}\hat{x}+\sin{\alpha}\hat{y}$                             & $\frac{U}{Q}$ & $\frac{V}{I}$ & \begin{tabular}[c]{@{}r@{}} Chiral sample or Half\\ waveplate oriented at\\ small angle with respect to\\ horizontal polarization\end{tabular}      
\end{tabular}
\end{ruledtabular}
\end{table*}
\par
The above formalism is a generalized one where all the different polarization anisotropy effects, namely, linear and circular diattenuation, and linear and circular retardance, are included in the framework.
Note that the expressions for experimentally measurable Stokes vectors corresponding to the Jones vectors presented above can easily be worked out using the standard algebra connecting Jones and Stokes formalism \cite{gupta2015wave}. In what follows, we complement the above weak measurement formalism with its corresponding classical field based analogue. We also provide a more detailed account of real and imaginary WVA of different polarization anisotropy effects and how they are manifested in the characteristic Stokes vector elements.
\subsection{Classical field based formalism}
\subsubsection{WVA of diattenuation}
\paragraph{Linear diattenuation ($x-y$)}
The real and imaginary WVA of linear diattenuation effect can be modeled using Eq.\eqref{eq4} and \eqref{eq5} using the generator $G_3^\mathfrak{a}$, and input state $\ket{\phi_i}=\frac{1}{\sqrt{2}}\begin{bmatrix}
    1&1
    \end{bmatrix}^T$ in the formalism described in the preceding sub-section. 
In the classical field analogue, this scenario can be mimicked through the interference of $+45^{o}$ polarized light in one path with slightly changed polarization state after experiencing a small linear diattenuation effect between horizontal ($x$) and vertical ($y$) polarization components in the other path. The corresponding electric fields are
\begin{equation}
    \pmb{E_1}=\mathfrak{\xi}\frac{\hat{x}+\hat{y}}{\sqrt{2}};\  \pmb{E_2}=\mathfrak{\xi}\frac{e^\alpha\hat{x}+e^{-\alpha}\hat{y}}{\sqrt{e^{2\alpha}+e^{-2\alpha}}}
    \label{eq8}
\end{equation}
$\mathfrak{\xi}$ is an arbitrary field amplitude factor. Note that $+\alpha$ and $-\alpha$ effect in both paths respectively (as mentioned in section\ref{sec:level2a}) is equivalent to $2\alpha$ effect in a single path. The \textit{imaginary} WVA can be obtained by a small phase offset $\epsilon_p$ from the exact destructive interference of $\pmb{E_1}$ and $\pmb{E_2}$ to yield the resultant field $\pmb{E}$ as \cite{duck1989sense,ritchie1991realization}
\begin{equation}
    \pmb{E}=(\cos{\epsilon_p}\pm i\sin{\epsilon_p})\pmb{E_1}-(cos{\epsilon_p}\mp i\sin{\epsilon_p})\pmb{E_2}
    \label{eq9}
\end{equation}
Eq.\eqref{eq8} and \eqref{eq9} can be used to obtain the expression for the corresponding Stokes vector elements ($S=[I\ Q\ U\ V]^T$) \cite{collett1993polarized,gupta2015wave}. In the weak coupling limit ($\alpha\rightarrow0$) \cite{aharonov1988result,duck1989sense,kofman2012nonperturbative,dressel2014colloquium}, the circular (elliptical) polarization descriptor 4th Stokes vector element ($\frac{V}{I}$) \cite{collett1993polarized,gupta2015wave} can be shown to exhibit WVA with decreasing $\epsilon_p$  as
\begin{equation}
    \frac{V}{I}\approx\mp\alpha\cot{\epsilon_p}
    \label{eq10}
\end{equation}
The corresponding expressions for \textit{real} WVA can be obtained by destructive interference of $\pmb{E_1}$ and $\pmb{E_2}$ with small amplitude offset $\epsilon_a$ ($=\tan^{-1}{\frac{1-a}{1+a}}$) to yield \cite{duck1989sense,ritchie1991realization}
\begin{equation}
    \pmb{E}=(\cos{\epsilon_a}\pm \sin{\epsilon_a})\pmb{E_1}-(cos{\epsilon_a}\mp \sin{\epsilon_a})\pmb{E_2}
    \label{eq11}
\end{equation}
Once again, in the weak coupling limit ($\alpha\rightarrow0$), the real WVA is manifested in the linear polarization descriptor Stokes vector element ($\frac{Q}{I}$) (for $\epsilon_a\rightarrow0$) as
\begin{equation}
    \frac{Q}{I}\approx\mp\alpha\cot{\epsilon_a}
    \label{eq12}
\end{equation}
Complete expressions for the relevant Stokes vector elements ($V,I$ for imaginary WVA and $Q,I$ for real WVA) are provided in Appendix\ref{appendix}. The electric fields and the polarization ellipse corresponding to WVA of linear diattenuation are pictorially illustrated in Fig.\ref{figure1}b. For real WVA, with varying $\epsilon_a$, the Stokes polarization state evolves in the Poincare sphere along the geodesic trajectory connecting the states of the two paths of the interferometer (the input state and the state after encountering the weak anisotropy effect). For imaginary WVA, the corresponding trajectory with varying $\epsilon_p$ lies in a plane that is perpendicular to the geodesic trajectory (Fig.\ref{figure1}b, iii.). This appears to be a general rule for all the anisotropy effects.
\paragraph{Circular diattenuation}
The real and imaginary WVA of this effect can be modeled using Eq.\eqref{eq4} and \eqref{eq5} with the generator matrix $G_2^\mathfrak{a}$, and with the choice of the input state $\ket{\phi_i}=\begin{bmatrix}
    1&0
    \end{bmatrix}^T$ in the weak value formalism. In the classical field based formalism, the corresponding expressions for the fields $\pmb{E_1}$ and $\pmb{E_2}$ for input horizontal ($x$) polarization are provided in Table\ref{table1}. This leads to \textit{imaginary} and \textit{real} WVA (respectively) of circular diattenuation in the Stokes vector elements as
 \begin{subequations}
    \begin{equation}
        \frac{U}{I}\approx\alpha\cot{\epsilon_p}
    \end{equation}
    \begin{equation}
    \frac{V}{I}\approx\alpha\cot{\epsilon_a}
     \end{equation}
     \label{eq13}
    \end{subequations}
The exact expressions for the above Stokes vector elements are provided in Appendix\ref{appendix}. The corresponding polarization state trajectories in the Poincare sphere for varying $\epsilon_{a\slash p}$ are shown in Fig.\ref{figure1}c.
\subsubsection{WVA of linear and circular retardance}
The above framework for WVA can be generalized for retardance properties also (shown in Table\ref{table1}). Full expressions for the relevant Stokes vector elements are provided in Appendix\ref{appendix}. From a fundamental point of view, it is interesting to note that the real and the imaginary WVAs of a given polarization anisotropy effect are manifested in different Stokes Vector elements ($Q,U,V$), which act as conjugate variables. It is interesting to note the quantum mechanical equivalence where the corresponding quantum Stokes operators satisfy commutation relations and the respective variances are also restricted by uncertainty relations \cite{bowen2002experimental}. From a practical point of view, manifestation of real and imaginary WVAs of  various polarization anisotropy effects in characteristic Stokes vector elements open up  the possibility of a weak value polarimeter for 
amplification and quantification of all the polarization anisotropy effects using a single experimental embodiment with relative experimental ease. We experimentally demonstrate this subsequently. 
\begin{figure*}
\includegraphics[width=150mm]{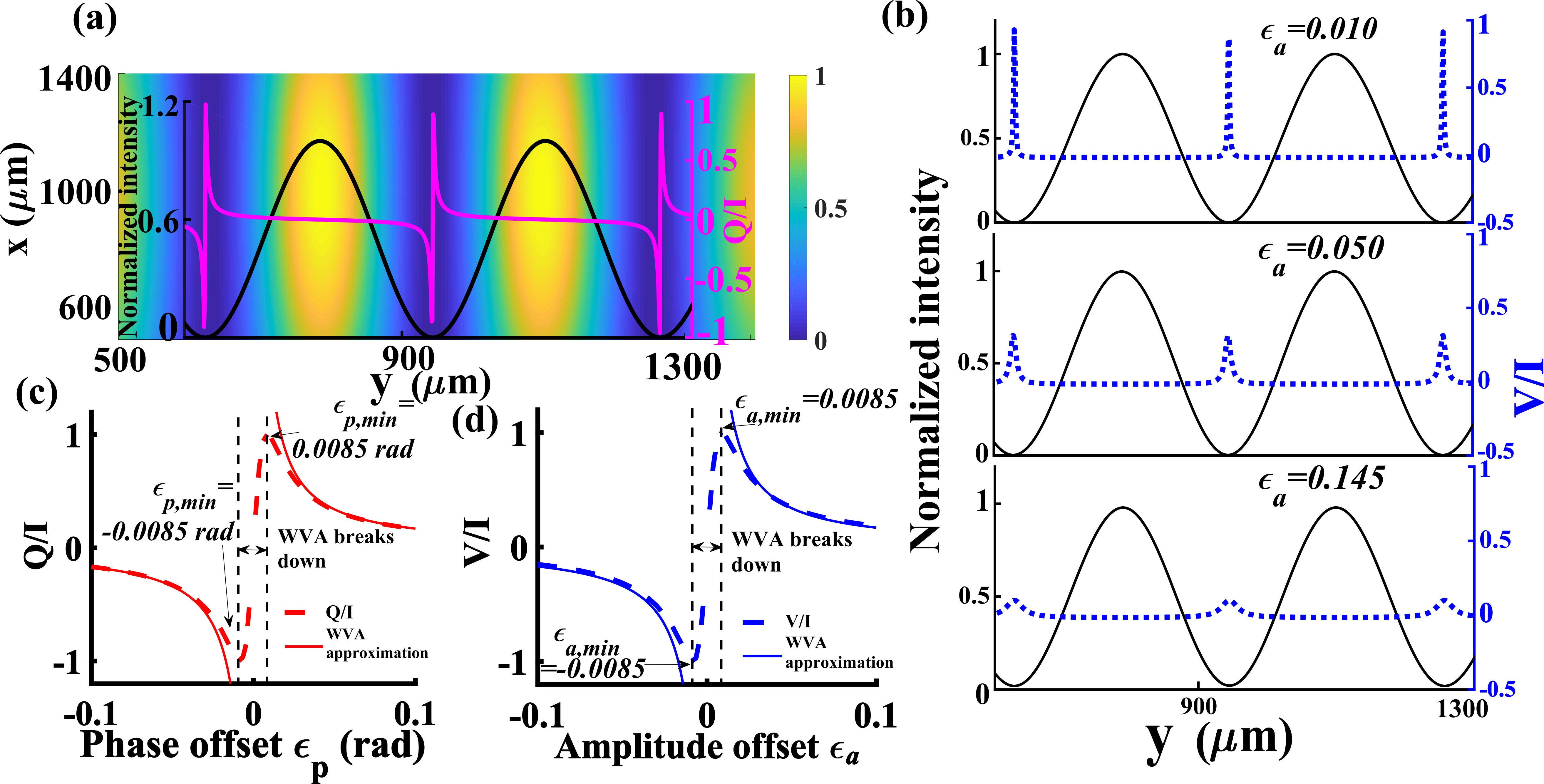}
\caption{\textit{Simulation of real and imaginary WVA} of linear retardance ($\alpha=0.017rad$). (a) Typical fringe profile of the interference of $+45^{o}$ polarized Gaussian beam (width = 1mm) with another beam having a slightly changed polarization state due to the weak linear retardance effect. The intensity profile (left axis, black line) and the spatial variation of $\frac{Q}{I}$ Stokes parameter (right axis, magenta line) along $y$-axis of the fringe are displayed. Colour bar represents magnitude of intensity. The $\frac{Q}{I}$ parameter exhibits \textit{\textbf{imaginary}} WVA. (b) Corresponding results for \textit{\textbf{real}} WVA. The intensity profile (left axis, black solid line) and spatial variations of the $\frac{V}{I}$ Stokes parameter (right axis, blue dashed line) are shown for three different values of amplitude offset parameter $\epsilon_a=0.010,0.050,0.145$. (c) Variation of $\frac{Q}{I}$ with the phase offset parameter $\epsilon_p$ (red solid circles) and corresponding theoretical fit to imaginary WVA ($\sim\alpha\cot{\epsilon_p}$) (red dashed line). (d) Variation of $\frac{V}{I}$ with $\epsilon_a$ (blue solid circles) and the real WVA fit ($\sim\alpha\cot{\epsilon_a}$, blue dashed line). The region where the WVA approximation breaks down ($\epsilon_{a/p,min}\lesssim\frac{\alpha}{2}$)are also marked.}
\label{fig2}
\end{figure*}
 \section{\label{sec:level3}experimental realization of sample measuring weak value polarimeter}
In our experiment, the 632.8 nm line of a He-Ne laser is passed through a rotatable polarizer P1 and is used to seed the interferometer (see Fig.\ref{figure1}a). The small polarization anisotropy effect is introduced in one arm using standard polarization optical elements noted in the 4th coloumn of Table\ref{table1}. The light beams in the two arms with slightly different polarization states then interfere at the exit port and the resulting fringes are imaged into a CCD camera. The spatially resolved Stokes vector elements across the interference fringe are measured using a combination of quarter waveplate QWP and a linear polarizer P2 \cite{collett1993polarized,gupta2015wave}. Standard Stokes vector measurement procedure \cite{collett1993polarized,gupta2015wave} was adopted for this purpose by performing six linear and circular polarization intensity measurements using QWP and P2 respectively; $I_H$: 
P2 at $0^{o}$ and QWP at $0^{o}$, $I_V$: P2 at $90^{o}$ and QWP at $90^{o}$, $I_P$: P2 at $45^{o}$ and QWP at $45^{o}$, $I_M$: P2 at $135^{o}$ and QWP at $135^{o}$, $I_R$: P2 at $0^{o}$ and QWP at $45^{o}$, $I_L$: P2 at $0^{o}$ and QWP at $135^{o}$ \cite{collett1993polarized,gupta2015wave}. For the quantification of the \textit{imaginary} WVA, the intensities in the two arms are kept equal (amplitude ratio $a=1$) and the spatial variation of the relevant Stokes vector elements around the position of the destructive interference intensity minima (corresponding to phase $\pi$) are recorded. Here, a single set of measurement of the spatial variations of the relevant Stokes vector element is sufficient. This spatial variation is used to generate its variation as a function of the small phase offset parameter (phase shift of $\epsilon_p$ from $\pi$), which increases gradually as one moves away from the destructive interference point. Probing the \textit{real} WVA, on the other hand, involves multiple set of measurements of the spatial variation of the relevant Stokes vector elements. In order to quantify the \textit{real} WVA, the relative intensities (or amplitude ratio $a$) of light in the two arms are varied using a variable neutral density filter ND (see Fig.\ref{figure1}a). For each value of $a$, the measured Stokes polarization parameters corresponding to the spatial position of the destructive interference (intensity minima) are used to generate the variations of the relevant Stokes parameters with the small amplitude offset parameter $\epsilon_a$. Calibration procedure of our experimental system is discussed in Appendix\ref{cali}.

\section{\label{sec:level4}RESULTS and DISCUSSION}
First, we illustrate the WVA concept through simulation of the interference experiment taking small linear retardance ($\alpha=0.017 rad$) as the weak polarization anisotropy effect and the corresponding results are shown in Fig.\ref{fig2}. For the simulations of the interference fringes and the corresponding imaginary and the real WVA of linear retardance, electric fields $\pmb{E}$ corresponding to Eq.\eqref{eq9} and \eqref{eq11} (respectively) are used by taking expressions for $\pmb{E_1}$ and $\pmb{E_2}$ from the fourth row of Table\ref{table1}. The  amplitude factor $\mathfrak{\xi}$ was taken to be a Gaussian (width=1mm) to mimic the light beam for the simulation of the interference fringes. The Stokes vector elements corresponding to these interfering fields were generated using standard Stokes algebra \cite{collett1993polarized,gupta2015wave} and the corresponding expressions are provided in Eq.\eqref{eqa9}, \eqref{eqa10} and Eq. \eqref{eqa11}, \eqref{eqa12} of Appendix\ref{appendix} for real and imaginary WVA, respectively.  As apparent from Fig.\ref{fig2}a, the  $\frac{Q}{I}$ Stokes parameter exhibits prominent enhancement at close vicinity of the intensity minima of destructive interference position. Accordingly, $\frac{Q}{I}$  increases rapidly as $\propto\alpha\cot{\epsilon_p}$ with decreasing small phase offset parameter $\epsilon_p$ (Fig.\ref{fig2}c), implying the manifestation of \textit{imaginary} WVA of linear retardance $\alpha$. The corresponding \textit{real} WVA of $\alpha$ is manifested in the $\frac{V}{I}$ Stokes parameter (see Fig.\ref{fig2}b and Fig.\ref{fig2}d). The  variation of $\frac{V}{I}$ parameter across the fringe are shown for three different values of small amplitude offset $\epsilon_a$ in Fig.\ref{fig2}b. The $\frac{V}{I}$ parameter also increases rapidly with decreasing $\epsilon_a$  and varies as $\propto\alpha\cot{\epsilon_a}$  (Fig.\ref{fig2}d) implying real WVA. However, as opposed to the approximate expressions, the exact magnitudes of WVA (obtained using Eq.\eqref{eqa9}, \eqref{eqa10} and Eq. \eqref{eqa11}, \eqref{eqa12} of Appendix\ref{appendix}) saturates and start decreasing below a certain value of the $\epsilon_{p/a}$ parameters ($\epsilon<\epsilon_{min}$), which is a universal characteristics of WVA \cite{aharonov1988result,duck1989sense,kofman2012nonperturbative,dressel2014colloquium}. In accordance with this and as evident from Fig.\ref{fig2}c and \ref{fig2}d, there is a limiting value of amplification and a corresponding minimum value of $\epsilon_{p\slash a}$  ($\epsilon_{min}\sim\frac{\alpha}{2}$) like in the conventional WVA. Here, the limits are set by the fundamental limit of degree of polarization ($\leq1$). These limiting values of the $\epsilon_{p/a}$ parameters in both the regions $\epsilon_{p\slash a}>0$ and $\epsilon_{p\slash a}<0$ for imaginary and real WVA are marked in  Fig.\ref{fig2}c and \ref{fig2}d, respectively. As $\epsilon_{p\slash a}$ approaches zero, the WVA approximation breaks down and the shift of the pointer profile also approaches zero, illustrating the non-diverging nature of WVA (Fig.\ref{fig2}c and \ref{fig2}d) \cite{aharonov1988result,duck1989sense,kofman2012nonperturbative,dressel2014colloquium}.
\begin{figure}
\includegraphics[width=75mm]{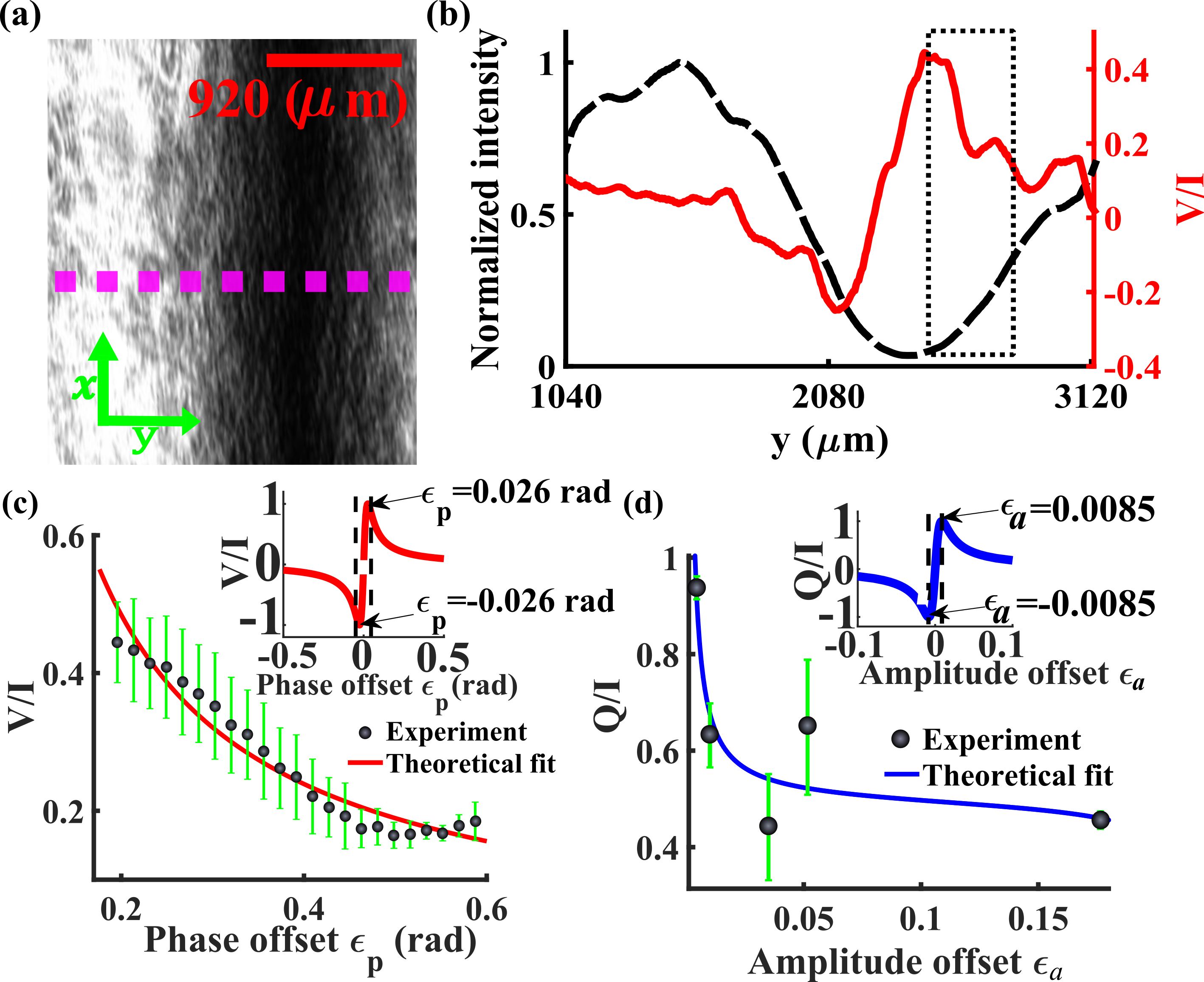}
\caption{\textit{Experimental results of real and imaginary WVA of linear diattenuation}. (a) Recorded fringe profile of the interference of $+45^{o}$ polarized Gaussian beam with another beam having $42^{o}$ polarization state due to the weak linear diattenuation effect $\alpha=0.051$. (b) The intensity profile (left axis, \textcolor{blue}{black dashed line}) and spatial variation of $\frac{V}{I}$ Stokes parameter (right axis, red solid line) along the magenta dashed line mark in (a). Rapid variation of $\frac{V}{I}$ around the destructive interference position, shown in black dotted box, is a manifestation of \textit{imaginary} WVA. (c) Variation of $\frac{V}{I}$ as a function of $\epsilon_p$ (black balls) and theoretical fit to imaginary WVA ($\alpha\cot{\epsilon_p}$ ) (red line). For real WVA, we took $\alpha=0.017$ by placing the linear polarizer at an angle $44^{o}$ w.r.to the horizontal axis in one arm of the interferometer. (d) Variation of $\frac{Q}{I}$ as a function of $\epsilon_a$ (black balls) and the corresponding \textit{real} WVA fit ($\alpha\cot{\epsilon_a}$, blue line). Error bars in (c) and (d) represent standard deviations. Corresponding theoretical predictions of imaginary and real WVA along with the maximum enhancement limit (minimum $\epsilon_{p\slash a}\ \text{i.e.},\ \epsilon_{min}$) are depicted in the inset of (c) and (d) respectively.}
\label{fig3}
\end{figure}
\par
The experimental results for linear ($x-y$) diattenuation (Fig.\ref{fig3}) reveal the role of Stokes parameters  $\frac{V}{I}$ and $\frac{Q}{I}$ in the imaginary and the real WVA of linear diattenuation  respectively as opposite to the case of linear retardance. Accordingly, the $\frac{V}{I}$ parameter approaches its maximum value near the destructive interference intensity minima (see Fig.\ref{fig3}b). The variation of $\frac{V}{I}$  with $\epsilon_p$ (Fig.\ref{fig3}c) shows a good agreement with the corresponding \textit{imaginary} WVA of linear ($x-y$) diattenuation ($\alpha=0.051$) as predicted in Eq.\eqref{eq10} ($\propto\alpha\cot{\epsilon_p}$). Similarly, $\frac{Q}{I}$ parameter also varies as ($\propto\alpha\cot{\epsilon_a}$) (see Fig.\ref{fig3}d) as predicted by Eq.\eqref{eq12} for \textit{real} WVA of linear diattenuation ($\alpha=0.017$). 
While the imaginary WVA results (Fig.\ref{fig3}c) are extracted from a single spatial map of $\frac{V}{I}$, the real WVA results (Fig.\ref{fig3}d) are derived from multiple measurements of $\frac{Q}{I}$ for varying intensity ratios at the two arms. This makes the latter more prone to experimental errors. Theoretical predictions (using Eq.\eqref{eqa1}, \eqref{eqa2}, \eqref{eqa3}, and \eqref{eqa4} of the Appendix\ref{appendix}) of imaginary and real WVA of linear ($x-y$) diattenuation are shown in the insets of Fig.\ref{fig3}c and \ref{fig3}d respectively. Once again, the limiting behaviour of the WVA are apparent when $\epsilon_{p\slash a}$ approaches the limit ($\epsilon_{min}\sim\frac{\alpha}{2}$). In the experiments, the values for $\epsilon_{p\slash a}$ was always greater than this limiting value $\epsilon_{min}$. The calibration procedure of the experimental system to extract the imaginary and real weak value (presented above) is discussed in Appendix\ref{cali}.
\par 
Fig.\ref{fig4} provides experimental results of WVA of circular diattenuation ($\alpha=0.012$). Imaginary WVA is manifested as increase of the  $\frac{U}{I}$  Stokes parameter with decreasing $\epsilon_p$ (Fig.\ref{fig4}a) showing good agreement with the corresponding prediction ($\propto\alpha\cot{\epsilon_p}$, Eq.\eqref{eq13}a). Real WVA (Fig.\ref{fig4}b) is manifested as ($\propto\alpha\cot{\epsilon_a}$) variation of the $\frac{V}{I}$ Stokes parameter as predicted by Eq.\eqref{eq13}b. Experiments were also performed for WVA of small optical rotation effect \cite{guchhait2020natural}. The results confirmed imaginary and real WVA of optical rotation $\alpha$, which were reflected in ($\propto\alpha\cot{\epsilon_{p\slash a}}$) variation of the $\frac{V}{I}$ and $\frac{U}{Q}$ Stokes parameters, respectively (not shown here) \cite{guchhait2020natural}. The above results demonstrate real and imaginary WVA of all the polarization anisotropy effects.
\begin{figure}
\includegraphics[width=75mm]{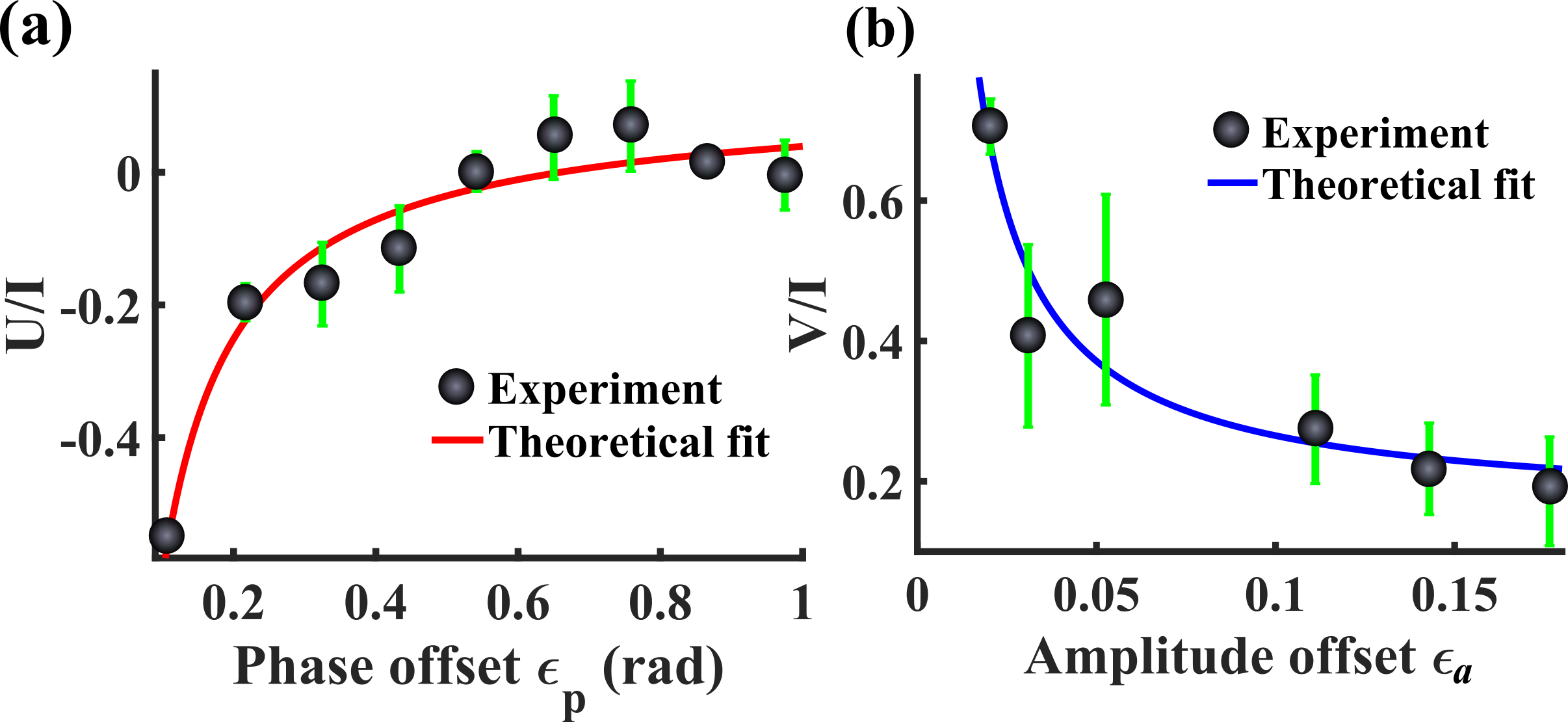}
\caption{\textit{Experimental results of real and imaginary WVA of circular diattenuation ($\alpha=0.012$)}. (a) Variation of $\frac{U}{I}$ Stokes parameter with $\epsilon_p$ (black balls) and corresponding theoretical fit to \textit{imaginary} WVA ($\alpha\cot{\epsilon_p}$) (red line). (b) Variation of $\frac{V}{I}$ as a function of $\epsilon_a$ (black balls) and the corresponding \textit{real} WVA fit ($\alpha\cot{\epsilon_a}$, blue line). Error bars represent standard deviations.}
\label{fig4}
\end{figure}
\par
As we proceed further, it is worth noting that even though at first sight it may appear that WVA of different polarization anisotropy effects are sometime manifested in the same Stokes vector elements, multiple polarimetry effects are in fact perfectly discernible through real and imaginary weak value measurement procedure and by clever choice of the input polarization states, as evident from Table\ref{table1}. 
 It is also pertinent to note that when multiple polarization effects are exhibited, extraction and quantification of individual  anisotropy parameters using conventional methods get confounded and necessitates cumbersome measurements such as the full $4\times4$ Mueller matrix measurements and its inverse analysis models that are often based on various assumptions and conditions. This interferometric WVA approach is potentially advantageous in this regard as it reduces the number of measurements and also obviates the need for the use of empirical inverse models for polarimetric quantification \cite{azzam2016stokes}.
\par
 The experimental WVA curves and their weak value fit for a given anisotropy effect provides the calibration curve for the quantification of any small anisotropy effect $\alpha$ of an unknown sample, which was validated by performing measurements for varying $\alpha$.(see details on calibration in Appendix\ref{cali}). We emphasize that in conventional polarimeters the sensitivity for quantification of $\alpha$ is typically of the order or lower than the sensitivity of individual Stokes vector or Mueller matrix elements \cite{collett1993polarized,gupta2015wave}. The most promising aspect of this interferometric WVA protocol, in this regard, is that one can achieve an amplification of the anisotropy parameter ($\alpha$) by a large WVA factor $\sim \cot{\epsilon}$. This essentially implies that, one can, in principle, quantify an anisotropy parameter ($\alpha$) from the measurement of $\sim \cot{\epsilon}$ times larger Stokes parameters ($\sim\alpha\cot{\epsilon}$). In other words, for a given sensitivity of a Stokes polarimeter, using this interferometric WVA scheme, one can extract and quantify the polarization anisotropy parameter which is $\sim\epsilon$ times smaller than the sensitivity of measurement of a given Stokes parameter ($Q/I,\ U/I,\ V/I$). Since the WVA parameter $\epsilon$ is a small parameter ($\epsilon\ll1$), this is equivalent to enhancement of the sensitivity of a conventional Stokes polarimeter, which is used in our WVA experiments. We would, however, like to note that akin to all other WVA approaches, here also the large amplification of polarization anisotropy parameter $ \alpha$ comes at the expense of the intensity signal, which may also limit polarimetric sensitivity \cite{ferrie2014weak,harris2017weak}. Thus, the actual benefit of this approach in terms of enhancement of sensitivity of a conventional Stokes polarimeter remains to be rigorously evaluated. Here, we provide initial evidence of sensitivity enhancement albeit for relatively moderate magnitude of anisotropy $\alpha$.
\par
\begin{figure}
\includegraphics[width=75mm]{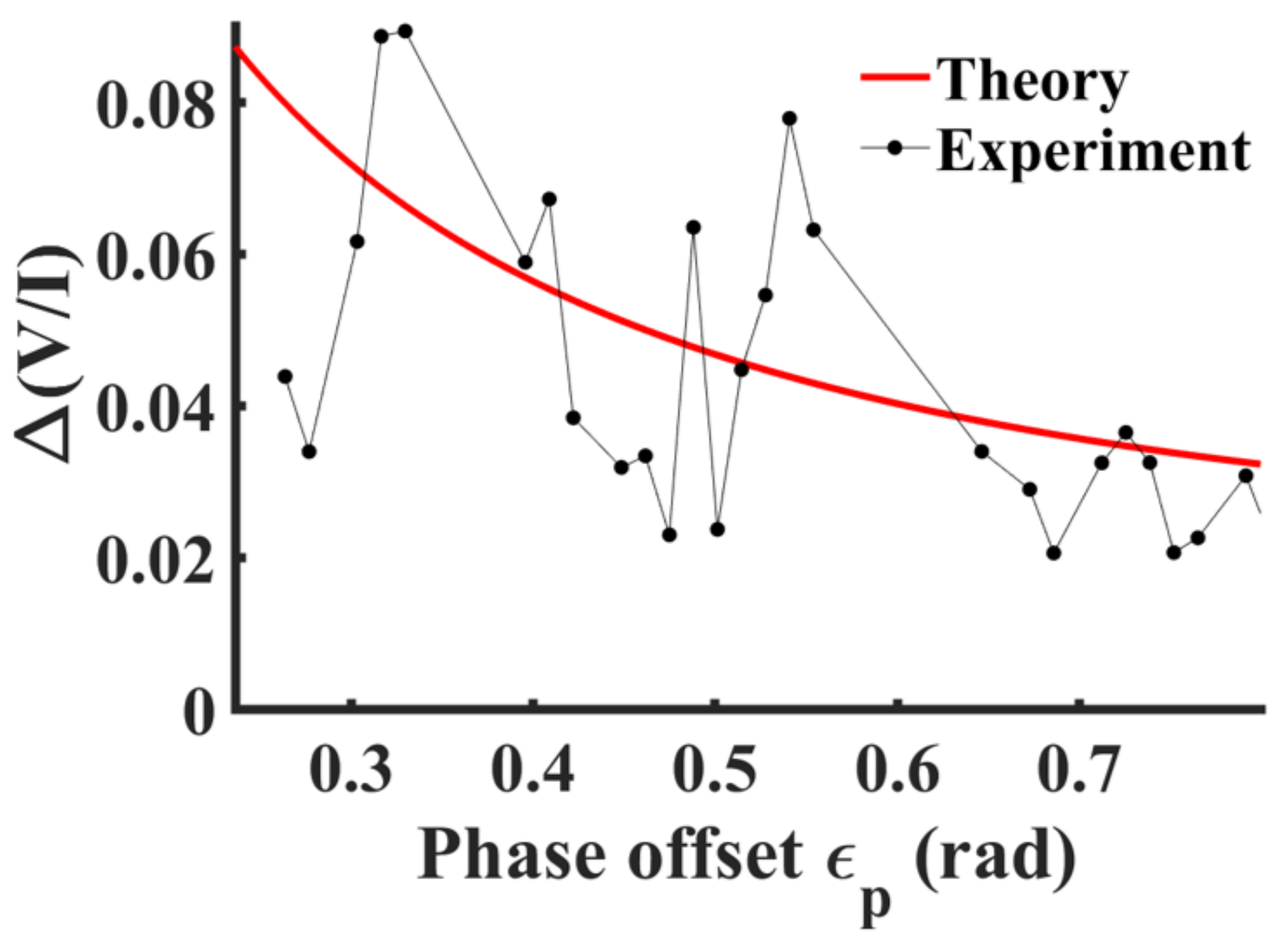}
\caption{\textit{Variation of the uncertainty in the intensity normalised Stokes parameter $\Delta(\frac{V}{I})$ with varying $\epsilon_p$ parameter. Here, $\Delta(\frac{V}{I})$ is an indicator of the the sensitivity of $V/I$. The results are shown for imaginary WVA of optical rotation ($\alpha=0.035$)}. Black solid circles depicts experimental sensitivity data and the corresponding theoretical prediction is shown by solid red line}
\label{fig5}
\end{figure}

 
For this purpose, we determine the sensitivity  of measurement of the intensity normalized Stokes polarization parameters ($\frac{S_i}{I};\ S_i=Q,\ U,\ V$) used in our interferometric WVA scheme, as this is relevant for the quantification of the polarization anisotropy parameter $\alpha$. The sensitivity here is limited by the corresponding intensity noise or uncertainty $\Delta(\frac{S_i}{I})$ 
(see Appendix \ref{sens} for the estimation of polarimetric sensitivity) \cite{perkins2010signal}. 
As an illustrative example, in Fig. \ref{fig5}, we have shown the variation of the uncertainty in the intensity normalised Stokes V parameter, 
$\Delta \frac{V}{I}$ with varying phase offset parameter $\epsilon_p$ for imaginary WVA of optical rotation. The results are shown for a value of $\alpha = 0.035rad$. The theoretical values (shown by red solid line) are generated using Eq.\eqref{eqnB1}.
For this purpose, 
Eq. \eqref{eqa15} and\eqref{eqa16} for imaginary WVA of optical rotation are used. The corresponding experimental values of $\Delta(\frac{V}{I})$ are shown by the black solid circles in Fig. \ref{fig5}. Fig. \ref{fig5} implies that as $\epsilon_p\rightarrow0$, sensitivity of determination of $(\frac{V}{I})$ deteriorates i.e., uncertainty in $\Delta(\frac{V}{I})$ increases which is expected, since overall intensity falls as $\epsilon_p$ approaches zero. 
As noted above, this is a generic feature of any WVA approach \cite{harris2017weak,ferrie2014weak}. It is pertinent to note here that WVA is not a null intensity detection technique as it uses near orthogonal pre-post selection of states corresponding to small overlap $\epsilon_{a\slash p}$ between the pre-post selected states \cite{aharonov1988result,duck1989sense}. Therefore, in WVA, there is an additional handle to optimize the amplification of any small physical parameter and the signal to noise ratio (SNR or sensitivity) by a judicial choice of the $\epsilon_{a\slash p}$ parameter. It has been earlier  demonstrated that due to this and other related advantages, in general WVA outperforms comparable standard strong measurements \cite{harris2017weak}. Our interferometric WVA scheme of polarimetry is no exception in this regard. As evident from Fig. \ref{fig5}, even though the sensitivity in determination of $(\frac{V}{I})$ deteriorates (uncertainty $\Delta(\frac{V}{I})$ increases) with decreasing $\epsilon_p$, the deterioration is not as drastic as the weak value amplification factor $\sim \cot{\epsilon_p}$. Thus, an optimal range of $\epsilon_{p}$ (0.2 to 0.8) can be be worked out where one can obtain significant amplification with an acceptable level of SNR. Using the sensitivity analysis of Appendix \ref{sens}, typical uncertainty (sensitivity) in the intensity normalized Stokes vector elements was determined to be $\sim 0.07$. Using this, the typical sensitivity in estimating the optical rotation, $\Delta \alpha$ was determined to be $0.008rad$ for the range of $\epsilon_{p}$ (0.2 to 0.8). Sensitivity of measurement of the same optical rotation ($\alpha \sim 0.035 rad$) using just the traditional Stokes polarization measurement was obtained to be $\sim 0.02 rad$. These  results provide evidence of the potential advantage and polarimetric sensitivity enhancement for the determination of optical rotation ($\alpha$) through the imaginary WVA of ($\frac{V}{I}$) by using the same Stokes polarization parameter measurement system in the interferometric WVA experimental embodiment.   
 \par
 \par
 \par We emphasize that this proof of concept of polarimetric sensitivity enhancement using the interferometric WVA approach is demonstrated using a dc Stokes polarimeter in our experimental configuration, which has inherently low sensitivity (we could not go below $\alpha\sim0.01 rad$ due to this limitation). The principle is however, expected to be valid for any conventional Stokes polarimeters irrespective of the specifics of the polarimeter. The initial results presented above on the enhancement of sensitivity are therefore quite promising and warrant further exploration towards detection of polarization anisotropy below the conventional limit through integration of this WVA scheme with relatively high sensitivity Stokes polarimeters \cite{azzam2016stokes}. 
\par
\section{\label{sec:level5}CONCLUSION}
In summary, using a simple yet profound philosophy of interferometric realization of weak value amplification of polarization anisotropy effect, we have introduced and experimentally demonstrated an interesting and useful concept of sample measuring weak value polarimeter. In this interferometric WVA approach, pre-post selection of states are obtained by near destructive interference of two paths of an interferometer with slightly different polarization states of light due to the presence of a weak anisotropy effect in one path. 
\par
This approach enables 
amplification and subsequent quantification of small magnitudes of all the sample polarization anisotropy effects in a single experimental embodiment. Like other WVA protocols, this amplification is obtained at the cost of the intensity of the output signal. But the WVA approach also provides a handle to optimize the amplification of polarization anisotropy effect and the polarimetric sensitivity by judiciously choosing the parameter $\epsilon_{a\slash p}$ that quantifies the overlap of pre-post selected states. Real and imaginary WVAs of a given sample anisotropy effect manifest themselves in different characteristic Stokes vector elements, which evolve in orthogonal trajectories in the Poincare sphere. This manifestation of WVA of polarization anisotropy in the conjugate Stokes parameters has an intriguing quantum mechanical equivalence in the commutation relations of the analogous quantum Stokes polarization operators and in their uncertainty relations \cite{bowen2002experimental}. This WVA protocol may thus bear useful and interesting consequences in the studies of non-trivial polarization properties of light in quantum theory and in non-classical polarization states. On practical grounds, the possibility of
significant enhancement of the sensitivity of polarization measurements using traditional Stokes polarimeters in the experimental embodiment of this WVA protocol is of general relevance  for studying wide range of weakly anisotropic materials and weak spin polarization optical effects for diverse applications \cite{gupta2015wave}. Finally, the extension of this proposed sample measuring weak value polarimeter concept to the other established domains of polarimetry like in spectroscopic and imaging polarimetry remains to be explored. We are currently expanding our investigations in these directions by integrating interferometric spectroscopy and imaging and microscopy methods \cite{huth2011infrared,hinz1998imaging} within the framework of weak value polarimeter.

\begin{acknowledgments}
We wish to acknowledge the support of the Indian Institute of Science Education and Research (IISER) Kolkata, an autonomous institute under the Ministry of Human Resource Development (MHRD), Govt. of India. We would like to acknowledge Science and Engineering  Research Board (SERB), Govt. of India for funding. Special thanks to Dr. Kamaraju Natarajan, for the experimental support.
\end{acknowledgments}

\appendix

\section{Manifestation of WVA of different polarization anisotropy effects in charactaeristic Stokes vector elements}
\label{appendix}
Here, we provide full expressions for the variations of the relevant Stokes vector elements with the small amplitude ($\epsilon_a$) and phase ($\epsilon_p$) offset parameters for real and imaginary (respectively) WVA of linear and circular diattenuation and retardance effects. We define four parameters as listed below
\begin{subequations}
    \begin{equation}
        p_1=\frac{1}{\sqrt{2}}+\frac{e^{\alpha}}{\sqrt{e^{2\alpha}+e^{-2\alpha}}}
    \end{equation}
    \begin{equation}
    p_2=\frac{1}{\sqrt{2}}-\frac{e^{\alpha}}{\sqrt{e^{2\alpha}+e^{-2\alpha}}}
     \end{equation}
     \begin{equation}
     p_3=\frac{1}{\sqrt{2}}-\frac{e^{-\alpha}}{\sqrt{e^{2\alpha}+e^{-2\alpha}}}
     \end{equation}
     \begin{equation}
     p_4=\frac{1}{\sqrt{2}}+\frac{e^{-\alpha}}{\sqrt{e^{2\alpha}+e^{-2\alpha}}}
      \end{equation}
    \end{subequations}
\subsection{Stokes parameters for linear diattenuation ($x-y$)}
As listed in Table\ref{table1} the real and imaginary WVA of small linear diattenuation ($x-y$) effect is manifested in the Stokes vector element $\frac{Q}{I}$ and $\frac{V}{I}$ respectively. The full expressions of the dependence of these Stokes parameters on $\epsilon_a$  and $\epsilon_p$ parameters are given as follows \cite{collett1993polarized,gupta2015wave}.
\subsubsection{Real WVA}
\begin{eqnarray}
    \label{eqa1}
    I&=&[\cos{\epsilon_a}p_2+\sin{\epsilon_a}p_1]^2+[\cos{\epsilon_a}p_3+\sin{\epsilon_a}p_4]^2\\
    \label{eqa2}
    Q&=&[\cos{\epsilon_a}p_2+\sin{\epsilon_a}p_1]^2-[\cos{\epsilon_a}p_3+\sin{\epsilon_a}p_4]^2
\end{eqnarray}
The expressions given in Eq.\eqref{eqa1} and \eqref{eqa2} in the limit $0<\epsilon_a\ll1$, lead to the WVA equation $\frac{Q}{I}\sim\alpha\cot{\epsilon_a}$ (Eq.\eqref{eq5}).
\subsubsection{Imaginary WVA}
Similarly, for imaginary WVA, the relevant Stokes parameters ($I,V$) are given as, respectively
\begin{eqnarray}
\label{eqa3}
    I&=&\cos^2{\epsilon_p}p_2^2
    +\sin^2{\epsilon_p}p_1^2
    +\cos^2{\epsilon_p}p_3^2
    +\sin^2{\epsilon_p}p_4^2\\
 \label{eqa4}
    V&=&2\cos{\epsilon_p}\sin{\epsilon_p}p_2p_4
    -2\cos{\epsilon_p}\sin{\epsilon_p}p_1p_3
\end{eqnarray}
Eq.\eqref{eqa3} and \eqref{eqa4} yields the familiar WVA equation as $\frac{V}{I}\sim\alpha\cot{\epsilon_p}$ in a straight forward way in the limit $0<\epsilon_p\ll1$.
\par
\subsection{Stokes parameters for circular diatenuation}
In a similar manner to the real and imaginary WVA of linear diattenuation ($x-y$), the electric fields corresponding to circular diattenuation effect (provided in Table.\ref{table1}) can be used to generate the Stokes parameters relevant to its WVA. The corresponding expressions are noted below \cite{collett1993polarized,gupta2015wave}.
\subsubsection{Real WVA}
\begin{eqnarray}
    \label{eqa5}
    I&=&[\cos{\epsilon_a}p_2+\sin{\epsilon_a}p_1]^2+[\cos{\epsilon_a}p_3+\sin{\epsilon_a}p_4]^2\\
    \label{eqa6}
    Q&=&[\cos{\epsilon_a}p_2+\sin{\epsilon_a}p_1]^2-[\cos{\epsilon_a}p_3+\sin{\epsilon_a}p_4]^2
\end{eqnarray}
\subsubsection{Imaginary WVA}
\begin{eqnarray}
    \label{eqa7}
    I&=&\cos^2{\epsilon_p}p_2^2
    +\sin^2{\epsilon_p}p_1^2
    +\cos^2{\epsilon_p}p_3^2
    +\sin^2{\epsilon_p}p_4^2\\
    \label{eqa8}
    U&=&2\cos{\epsilon_p}\sin{\epsilon_p}p_2p_4
    -2\cos{\epsilon_p}\sin{\epsilon_p}p_1p_3
\end{eqnarray}
\subsection{Stokes parameters relevant to linear retardance}
\subsubsection{Real WVA}
\begin{eqnarray}
\label{eqa9}
I&=&4(\sin^2{\frac{\alpha}{2}}\cos^2{\epsilon_a}+\cos^2{\frac{\alpha}{2}}\sin^2{\epsilon_a})\\
\label{eqa10}
V&=&4(\sin{\alpha}\cos{\alpha}\cos{\epsilon_a}\sin{\epsilon_a}\nonumber\\&-&\sin{\alpha}\sin^2{\frac{\alpha}{2}}\cos^2{\epsilon_a}\nonumber\\&-&\sin{\alpha}\cos^2{\frac{\alpha}{2}}\sin^2{\epsilon_a})
\end{eqnarray}
\subsubsection{Imaginary WVA}
\begin{eqnarray}
\label{eqa11}
I&=&2(1-\cos{2\epsilon_p}\cos{\alpha})\\
\label{eqa12}
Q&=&2\sin{2\epsilon_p}\sin{\alpha}
\end{eqnarray}
\subsection{Stokes parameters relevant to circular retardance}
\subsubsection{Real WVA}
\begin{eqnarray}
\label{eqa13}
Q&=&(\cos{\epsilon_a}+\sin{\epsilon_a})^2+(\cos{\epsilon_a}-\sin{\epsilon_a})^2\cos{2\alpha}\nonumber\\
&-&2\cos{\alpha}\cos{2\epsilon_a}\\
\label{eqa14}
U&=&\sin{2\alpha}(\cos{\epsilon_a}-\sin{\epsilon_a})^2-2\sin{\alpha}\cos{2\epsilon_a}
\end{eqnarray}
\subsubsection{Imaginary WVA}
\begin{eqnarray}
\label{eqa15}
I&=&2(1-\cos{\alpha}\cos{2\epsilon_p})\\
\label{eqa16}
V&=&2\sin{\alpha}\sin{2\epsilon_p}
\end{eqnarray}
\section{Calibration of the experimental system}
\label{cali}
Two different sets of calibrations are involved in this interferometric WVA scheme: (i) Calibration of the interferometric WVA parameters, namely, the phase offset ($\epsilon_p$) and the amplitude offset ($\epsilon_a$) parameters in the interferometric set-up, (ii) calibration of the WVA curve of the different characteristic Stokes vector elements encoding information on the different polarization anisotropy effects. Here, we briefly outline the procedure.
\begin{enumerate}[wide, labelwidth=!, labelindent=0pt]
\item For generating the experimental imaginary WVA curve, the values for the phase offset parameter $\epsilon_p$ at different spatial positions in the CCD image plane are calculated from the pixel-wise spatial variation of the phase of the interference fringe. For this purpose, the phase difference of $2\pi$ between two consecutive destructive spatial positions of the fringe profile on the CCD plane was marked and the pixel corresponding to the destructive interference point was assigned a value of $\epsilon_p$ =0. The values of $\epsilon_p$ corresponding to consecutive CCD pixels away from the destructive interference point was subsequently obtained. For the given pixel dimension of $3.45\mu m$, the maximum resolution of the phase offset parameter (the minimum value of $\epsilon_p$ that can be used in the WVA scheme) in our experimental system was determined to be $\Delta\epsilon_p\sim0.0132 rad$.   
The flatness in the spatial variation (or variation as a function of $\epsilon_p$) of the different characteristic Stokes vector elements in the absence of anisotropy effects ($\alpha=0$) were ensured by performing blank (without sample) measurements. For generating the experimental real WVA curve, the amplitude offset parameter $\epsilon_a$ is determined by taking the ratio of intensity in the two arms of the interferometer. If the intensity is equal in both arms then, $\epsilon_a = 0$. Tuning the ND (see Fig.\ref{figure1}), we varied the intensity in one of the arms and determined $\epsilon_a$ in each case. For the given sensitivity of $\sim0.001 \mu W$ of the detector used, the minimum value of the amplitude offset parameter in our experimental system is found to be $\Delta\epsilon_a\sim0.01$ . Once again, the flatness of the variation of the Stokes parameters with varying $\epsilon_a$ in absence of the anisotropy effects was ensured by performing blank measurements.  

\item The variations of the different characteristic Stokes vector elements with varying $\epsilon_p/a$ parameters for imaginary and real WVAs (respectively) were generally fitted with the function $A\alpha\cot{\epsilon_{p/a}}+B$. Note that in ideal case, the parameters $A$ and $B$ are expected to be unity and zero respectively. These parameters were therefore determined by performing measurements on known magnitudes of the different polarization anisotropy effects ($\alpha$). The typical values for the $A$ and $B$ parameters of our experimental system were found to be in the range $0.80$ to $1.08$, and $0.002$ to $0.010$, respectively. Several factors contribute to these variations, e.g., due to the non-ideal polarization optical components (polarizers and waveplates), slight misalignment and other experimental imperfections etc. Thus, the above-mentioned WVA fits (with known $A$ and $B$ experimental parameters for respective anisotropy effects) provide the calibration WVA curve of the experimental system for measurement of any unknown anisotropy parameter $\alpha$. The linearity of the values of $\alpha$ estimated from the WVA fitting, with varying magnitude of the input anisotropy parameter was also confirmed by performing multiple measurements.
\end{enumerate}

\section{Sensitivity estimation}
\label{sens}
In the context of quantification of polarization anisotropy $\alpha$, we are interested in the sensitivity of determining the intensity normalized Stokes parameters ($\frac{S_i}{I};\ S_i=Q,\ U,\ V$, $I$ is the total intensity, the sum of intensities of any two orthogonal polarizations) The sensitivity  of measurement of ($\frac{S_i}{I}$) used in our interferometric WVA scheme is limited by the noise or uncertainty $\Delta(\frac{S_i}{I})$ i.e., smaller $\Delta(\frac{S_i}{I})$ corresponds to better sensitivity. In this regard, the signal to noise ratio (SNR) of the normalized Stokes parameters can be formally defined as $SNR(\frac{S_i}{I})=\frac{\frac{S_i}{I}}{\Delta(\frac{S_i}{I})}$. Here, the uncertainty $\Delta(\frac{S_i}{I})$ primarily arises due to the photon shot noise and the readout noise of the detector. For un-normalized Stokes parameter $S$, the corresponding noise or uncertainty is $\Delta S_i\sim \sqrt{I+2R^2}$ where $R$ is the readout noise \cite{perkins2010signal}. The readout noise of our experimental detector was determined from the recorded images in the camera in dark condition and the value was estimated to be $~0.2$. Thus, in our case, the overall noise was mainly shot noise limited, i.e., $\Delta S_i\sim \sqrt{I}$. Based on these definitions and by using standard error propagation method, the uncertainty 
of the normalized Stokes parameters $\Delta(\frac{S_i}{I})$ can be obtained as \cite{bevington2003data}.



\begin{equation}
    \Delta\bigg({\frac{S_i}{I}}\bigg)= \frac{\frac{S_i}{I}}{SNR(\frac{S_i}{I})} = \frac{\Delta S_i (I+S_i)}{I^2}
     \label{eqnB1}
\end{equation}

The above equation was used to determine the sensitivity of normalized Stokes parameter measurements $\Delta(\frac{S_i}{I})$ and the resultant sensitivity of determination of all the sample polarization anisotriopy effects $\Delta \alpha$ of our interferometric WVA polarimeter. Illustrative results of sensitivity analysis for imaginary WVA of optical rotation are presented in Fig. \ref{fig5}.

\bibliography{apssamp}

\end{document}